\documentclass[useAMS,usenatbib,usegraphicx]{mn2e}

\title[Spatial distribution in $\sigma$~Orionis]{Spatial distribution of stars
and brown dwarfs in $\sigma$ Orionis}     
\author[Jos\'e A. Caballero]{Jos\'e Antonio Caballero$^{1,2}$\\
$^{1}$Max-Planck-Institut f\"ur Astronomie, K\"onigstuhl 17, D-69117
Heidelberg, Germany \\ 
$^{2}$Dpto. de Astrof\'{\i}sica y Ciencias de la Atm\'osfera, Facultad de
Ciencias F\'{\i}sicas, Universidad Complutense de Madrid, \\ 
E-28040 Madrid, Spain. E-mail: caballero@astrax.fis.ucm.es}
\begin{document}

\date{Accepted 2007 October --. Received 2007 October --; 
in original form 2007 August 13}

\pagerange{\pageref{firstpage}--\pageref{lastpage}} \pubyear{2007}

\maketitle

\label{firstpage}

\begin{abstract}
I~have re-visited the spatial distribution of stars and high-mass brown
dwarfs in the $\sigma$~Orionis cluster ($\sim$3\,Ma, $\sim$360\,pc). 
The input was a catalogue of 340 cluster members and candidates at separations
less than 30\,arcmin to $\sigma$~Ori~AB.
Of them, 70\,\% have features of extreme youth.
I~fitted the normalised cumulative number of objects counting from the cluster
centre to several power-law, exponential and King radial distributions.
The cluster seems to have two components: a dense core that extends from the
centre to $r \approx$ 20\,arcmin and a rarified halo at larger separations.
The radial distribution in the core follows a power-law proportional to $r^1$,
which corresponds to a volume density proportional to $r^{-2}$.
This is consistent with the collapse of an isothermal spherical molecular
cloud.  
The stars more massive than 3.7\,$M_\odot$ concentrate, however, towards the
cluster centre, where there is also an apparent deficit of very low-mass objects
($M <$ 0.16\,$M_\odot$). 
Last, I~demonstrated through Monte Carlo simulations that the cluster is
azimuthally asymmetric, with a filamentary overdensity of objects that runs from
the cluster centre to the Horsehead~Nebula.
\end{abstract}


\section{Introduction}

The $\sigma$~Orionis region in the {Ori~OB~1~b} association is finally
becoming recognised as one of the most important young open clusters, with an
age of only about 3\,Ma. 
In the discovery paper, Garrison (1967) used the term ``clustering'' to refer to
an agglomeration of fifteen B-type stars surrounding and including the multiple
star {$\sigma$~Ori}. 
Afterwards, Lyng\aa~(1981) tabulated $\sigma$~Orionis in his catalogue of open
clusters.
Since the rediscovery of the cluster by Wolk (1996) and its subsequent study in
depth, which has revealed the most numerous and best known substellar
population (B\'ejar et~al. 1999; Zapatero Osorio et~al. 2000, 2002; Caballero
et~al. 2007), only a few authors have investigated the $\sigma$~Orionis spatial
distribution.
In particular, B\'ejar et~al. (2004) and Sherry et~al. (2004) analysed the
radial distribution of $\sigma$~Orionis cluster members and candidates in annuli
of width $\Delta r$ as a function of the separation $r$ to $\sigma$~Ori~AB.
To maximise the number of objects per annulus and minimise the Poissonian
errors, $\Delta r$ must be wide. 
This leads to have few annuli (no more than 12 in the $r$ = 0--30\,arcmin
interval) to fit to a suitable radial profile (exponential decay -- B\'ejar
et~al. 2004; King -- Sherry et~al.~2004). 
Both studies agree that the cluster may extend only up to $\sim$25--30\,arcmin.
The low surface density of cluster members at larger separations, the sharp
increase of extinction due to the nearby {Horsehead Nebula}-{Flame
Nebula}-{IC~434} complex and the closeness to (or even overlapping with)
other stellar populations in the Orion Belt surrounding {Alnitak}
($\zeta$~Ori) and {Alnilam} ($\epsilon$~Ori) prevent from suitably
broaden the radial distribution analysis (Caballero 2007a).
At the canonical heliocentric distance to $\sigma$~Orionis of 360\,pc (e.g.
Brown, de~Geus \& de~Zeeuw 1994), the cluster would have an approximate radius
of~3\,pc. 
 
In spite of the agreement on the size of $\sigma$~Orionis, the fits and the
profiles in B\'ejar et~al. (2004) and Sherry et~al. (2004) seem to be rather
incomplete and inappropiate, respectively. 
On the one hand, the King models were designed for tidally truncated globular
clusters (King 1962, 1966; Meylan 1987), and have also been satisfactorily used
for describing galaxies (e.g. Kormendy 1977; Binggeli, Sandage \& Tarenghi
1984). 
These systems have had enough time to be isothermal, on the contrary to very
young open clusters like $\sigma$~Orionis, where only gravitational relaxation
by initial mixing may have occurred (King 1962). 
On the other hand, B\'ejar et~al. (2004) exclusively focused on the cluster
substellar population.
Besides, the exponential fit in B\'ejar et~al. (2004) only accounted for the
five innermost annuli, which leaded to a high uncertainty in the derived
parameters.
Last, in the works by Sherry et~al. (2004) and B\'ejar et~al. (2004), the
input list of cluster members and candidates came from $VRI/IZ$ optical surveys.
Many sources in both analysis had no near-infrared or spectroscopic follow-up. 

For a correct study of the spatial distribution in $\sigma$~Orionis, it is
therefore necessary to use new fitting radial profiles and an input catalogue as
comprehensive as possible.
It must cover a wide mass interval.
Maximum completeness and minimum contamination of the catalogue are also
desired. 
These requirements are verified by the {\em Mayrit} catalogue, which tabulates
339 $\sigma$~Orionis members and candidates in a 30\,arcmin-radius circular area
centred on $\sigma$~Ori~AB (Caballero 2007c). 
Of them, 241 display features of extreme youth (e.g. OB spectral types, Li~{\sc
i} in absorption, H$\alpha$ in strong emission, spectral signatures of low
gravity, near- and mid-infrared excesses due to discs). 
The catalogue covers three orders of magnitude in mass, from the
$\sim$18+12\,$M_\odot$ of the O9.5V+B0.5V binary $\sigma$~Ori~AB to the
$\sim$0.033\,$M_\odot$ of the brown dwarf {B05~2.03--617} (Caballero \&
Chabrier, in~prep.).
Accounting for $\sigma$~Ori~A and B as different objects separated by
$\sim$0.25\,arcsec, then the equatorial coordinates of 340 young stars, brown 
dwarfs and cluster member candidates are available.
I will use this input catalogue to investigate the radial and azimuthal
distribution of objects in the $\sigma$~Orionis~region.

\section{Analysis and results}

\subsection{Cluster centre and radial gradient}
\label{clustercentre}

\begin{figure}
\centering
\includegraphics[width=0.57\textwidth]{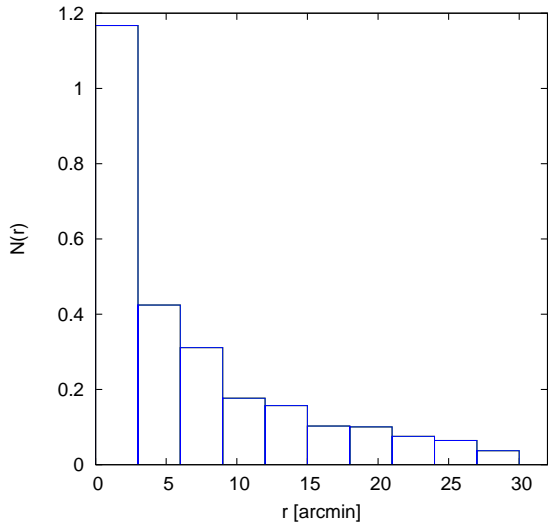}
\caption{Binned surface density profile for $\sigma$~Orionis with Caballero
(2007c) data. 
The mean number of objects per annulus is 34 for a width $\Delta r$ = 3\,arcmin.
All the figures are in colour in the online article on Synergy.} 
\label{orhist}
\end{figure}

Caballero (2007a) showed that $\sim$46\,\% of the mass in the
$\sigma$~Orionis stars with $M \ga 1.2\,M_\odot$ is contained in the
quintuple Trapezium-like system that gives the name to the cluster, and
$\sim$29\,\% only in the AB components. 
This suggests using the binary as the cluster centre ($r$ = 0).
From the masses for the objects in the {\em Mayrit} catalogue derived in
Caballero \& Chabrier (in~prep.), one third of the total cluster mass is
encircled in the innermost 5\,arcmin.
If the mass were homogenously distributed within the survey area, the
innermost 5\,arcmin would contain only 2.8\,\% of the total cluster
mass [$(5/30)^2 \approx 0.028$]. 
I~will consider $\sigma$~Ori~AB as the origin of coordinates because of: 
($i$) simplicity (the coordinates of the binary are well determined by {\em
Hipparcos}; the actual coordinates of the cluster barycentre may change when a
different input list of cluster members and individual masses is used);
($ii$) reflection of the geometry of the {\em Mayrit} survey in Caballero
(2007c), which was centred on $\sigma$~Ori~AB; and 
($iii$) uniformity with previous works (especially with B\'ejar et~al. 2004 and
Sherry et~al. 2004, who also used $\sigma$~Ori~AB as the coordinate origin).
There might be an additional reason:
the largest mass aggregation is probably associated to the
densest region of the original molecular cloud where the fragmentation and star 
formation initiated (assuming that the origin of the referece frame is
locked to the --moving-- cluster barycentre).
This reason may be unconvincing, because highly turbulent (and fractal-like?)
molecular clouds probably do not have a ``centre'' that can be defined in any
sensible way, as shown in the simulations of Bonnell, Bate \& Vine (2003).
See Section~\ref{azimuthalasymmetry} in this work and fig.~1 in Caballero
(2007a) for pictorical views of the spatial distribution of confirmed and
candidate cluster members in the $\sigma$\,Orionis region, and Caballero (2007b)
for a description of the cluster centre. 
The old-fashioned plot of the surface density is shown in
Fig.~\ref{orhist}; 
see similar plots in Sherry et~al. (2004) and B\'ejar et~al. (2004) for
comparison.

\begin{figure}
\centering
\includegraphics[width=0.57\textwidth]{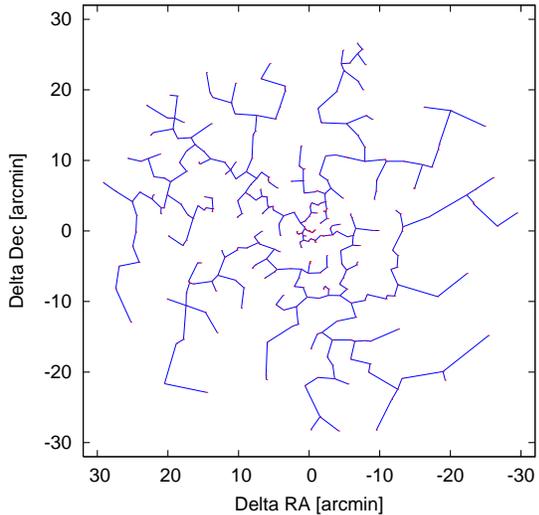}
\caption{Euclidean minimum spanning tree for $\sigma$~Orionis 
(${\mathcal Q} \approx$ 0.88).
I~have used a {\sc matlab/octave} function based on an open code by F.~W.~J. van
der Berg (University of Copenhagen), based in its turn on an algorithm by
Hillier and Lieberman (2001).
The minimum spanning trees for {$\rho$~Ophiuchus}, {IC~2391}, {IC~348}, {Taurus}
and {Chamaeleon} are shown in Cartwright \& Whitworth (2004).}
\label{omst}
\end{figure}

There exists a ${\mathcal Q}$-parameter which quickly, and simply, shows if
a distribution of cluster members is smooth (a large-scale radial density
gradient) or clumpy a (multiscale fractal subclustering; Cartwright \& Whitworth
2004).
The ${\mathcal Q}$-parameter is defined by: 

\begin{equation}
{\mathcal Q} = \frac{\overline{m}}{\overline{s}},
\end{equation}

\noindent where $m$ and $s$ are the edge length of the Euclidean minimum
spanning tree and the separation between cluster members, respectively.
A minimum spanning tree is a network (``graph'') of $N_{\rm max}$-1 lines
(``edges'') that connect the $N_{\rm max}$ objects (``nodes'') in the shortest
possible way under the condition no closed loops allowed.
See, e.g., Graham, Clowes \& Campusano (1995) for an application of minimum
spanning trees in Astrophysics\footnote{The first algorithm for finding a
minimum spanning tree was developed to find an efficient electrical coverage of
Czech Moravia (Bor\r{u}vka~1926)}.
The normalization factors are $(N_{\rm max} \pi r_{\rm max}^2)^{1/2} / (N_{\rm
max} - 1)$ and $r_{\rm max}$ for $m$ and $s$ ($N_{\rm max}$ is the total number
of cluster members and $\pi r_{\rm max}^2$ is the area of the circular survey).

For the 340 $\sigma$~Orionis cluster member and candidates, I~have measured
$\overline{m}$ = 0.589 and $\overline{s}$ = 0.668.
Therefore, the Cartwright \& Whitworth (2004) parameter is ${\mathcal Q}
\approx$ 0.88.
This value is larger than 0.80, which distinguish $\sigma$~Orionis as a cluster
with a smooth large-scale radial density gradient and a moderate degree of
central concentration.
This concentration is larger than in $\rho$~Ophiuchus, other cluster
with a radial density gradient, but less than in IC~348 (Cartwright \& Whitworth
2004). 
Other sparse clusters and star-forming regions, like IC~2391, Taurus and
Chamaeleon, have ${\mathcal Q}$-parameters in the interval 0.47--0.67, which
indicates that they have, on the contrary, subestructure with fractal dimensions
between 1.5 and~2.5.

\subsection{Surface density and cumulative number of objects}
\label{surfacedensity}

I present an innovative, accurate, simple method to derive the actual
expression of the surface density as a distance from the cluster centre,
$\sigma(r)$.
It can be applied to other open and globular clusters and galaxies.
The normalised cumulative number of objects counting from the cluster centre,
$f(r)$, is: 

\begin{equation}
f(r) = \frac{N(r)}{N(r_{\rm max})},
\end{equation}

\noindent where $N(r)$ is the total number of stars in projection within a
distance $r$ of the centre.
If there is azimuthal symmetry, $N(r)$ is related to the surface density
through the following expression: 

\begin{equation}
N(r) = 2 \pi \int_0^r dr' ~ r' ~ \sigma(r').
\end{equation}

\noindent 
The relatively high value of the ${\mathcal Q}$-parameter of
$\sigma$~Orionis supports the hypothesis of azimuthal symmetry in this cluster
in particular.
For systems without azimuthal symmetry (e.g. elliptical galaxies), use instead: 
\begin{equation}
N(r) = \int_0^r \int_0^{2 \pi} ~ dr' d\theta ~ 
r' ~ \sigma(r',\theta). 
\end{equation} 
\noindent 
The surface $\sigma(r)$ and volume $\rho(r)$ densities are linked through the
simple relation $\sigma(r) = 2 r \rho(r)$.
This equality comes from:

\begin{equation}
N(r) = 4 \pi \int_0^r dr' ~ r'^2 ~ \rho(r') = 2 \pi \int_0^r dr' ~ r' ~
\sigma(r') ~~~ [\forall r],
\end{equation}

\noindent assuming again azimuthal symmetry\footnote{Cartwright \&
Whitworth (2004) demonstrated that the equality is based on a fallacious
assumption. 
The differences between $\sigma(r)$ and $2 r \rho(r)$, linked to distribution
functions of type 2D1 and 3D2 in the nomenclature by those authors, are,
however, too small to be considered in this~work.}. 
The function $f(r)$ varies from 0 at $r$ = 0 to 1 at $r$ = $r_{\rm max}$.
In the {\em Mayrit} survey, $r_{\rm max}$ = 30\,arcmin and $N(r_{\rm max})$
($\equiv$ $N_{\rm max}$) = 340. 
In the discrete approximation, $N(r) \approx N^*(r)$ and:

\begin{equation}
f(r) \approx f^*(r) = 
\frac{\sum_{i=1}^{N^*(r)} i}{\sum_{i=1}^{N(r_{\rm max})} i} =
\frac{N^*(r)}{N_{\rm max}}.
\end{equation}

I~have investigated several functional expressions of $\sigma(r)$ that fit in
more or less degree the observed normalised cumulative number of objects,
$f^*(r)$. 
A general expression for a power-law surface density of index $\delta-2$ is:

\begin{equation}
\sigma(r,\delta) = \frac{\delta N_{\rm max} r^{\delta-2}}{2 \pi r_{\rm
max}^\delta},
\end{equation}

\noindent which, after integration, leads to a simple expression for~$f(r)$:

\begin{equation}
f(r,\delta) = \left( \frac{r}{r_{\rm max}} \right)^\delta.
\end{equation}

\noindent In this approach, the objects are uniformly distributed in a circular
area if $\delta = 2$ ($\sigma$ = constant).
Surface densities with parameter $\delta < 0$, that predict a lower number of
objects close to the centre, were obvioulsy not considered.

Following B\'ejar et~al. (2004), I~have also studied two expressions of
exponential decay of the surface density: 

\begin{eqnarray}
\sigma(r,\epsilon) = \sigma_0 e^{-\epsilon r} \\
\sigma_0 = \frac{N_{\rm max}}{2 \pi} 
\frac{1}{\frac{1}{\epsilon^2} - e^{-\epsilon r_{\rm max}} 
\left( \frac{1}{\epsilon^2} + \frac{r_{\rm max}}{\epsilon} \right)} \\
f(r) = \frac{1 - e^{-\epsilon r}}{1 - e^{-\epsilon r_{\rm max}}}
\end{eqnarray}

\noindent and:

\begin{eqnarray}
\sigma(r,\epsilon) = \sigma_0 e^{-\epsilon r^2} \\
\sigma_0 = \frac{\epsilon N_{\rm max}}{\pi} 
\frac{1}{1 - e^{-\epsilon r_{\rm max}^2}} \\
f(r) = \frac{1 - e^{-\epsilon r^2}}{1 - e^{-\epsilon r_{\rm max}^2}}.
\end{eqnarray}

Finally, I~have also investigated the King (1962) profile for gravitationally
relaxed globular clusters.
Close to the centre, the surface density can be expressed by:

\begin{equation}
\sigma(r) \approx \sigma_c(r) = \frac{\sigma_0}{1 + (r / r_c)^2},
\end{equation}

\noindent where $r_c$ is the core radius and $\sigma_0$ is the central surface
density.
In the limit of the cluster, the surface density is:

\begin{equation}
\sigma(r) \approx \sigma_t(r) = \sigma_1 \left( \frac{1}{r} - \frac{1}{r_t}
\right)^2, 
\end{equation}

\noindent where $r_t$ is the tidal radius (the value of $r$ at which
$\sigma_t(r)$ reaches zero) and $\sigma_1$ is a constant.
The overall normalised cumulative number of objects that embodies $\sigma_c(r)$
and $\sigma_t(r)$ is, following the nomenclature by King (1962):

\begin{equation}
f(r) = \frac
{\log{(1+x)} - 4 \frac{(1+x)^{1/2}-1}{(1+x_t)^{1/2}}+\frac{x}{1+x_t}}
{\log{(1+x_{\rm max})} - 4 \frac{(1+x_{\rm max})^{1/2}-1}{(1+x_t)^{1/2}}+\frac{x_{\rm max}}{1+x_t}},
\end{equation}

\noindent where $x = (r/r_c)^2$, $x_t = (r_t/r_c)^2$ and 
$x_{\rm max} = (r_{\rm max}/r_c)^2$. 

\begin{figure}
\centering
\includegraphics[width=0.57\textwidth]{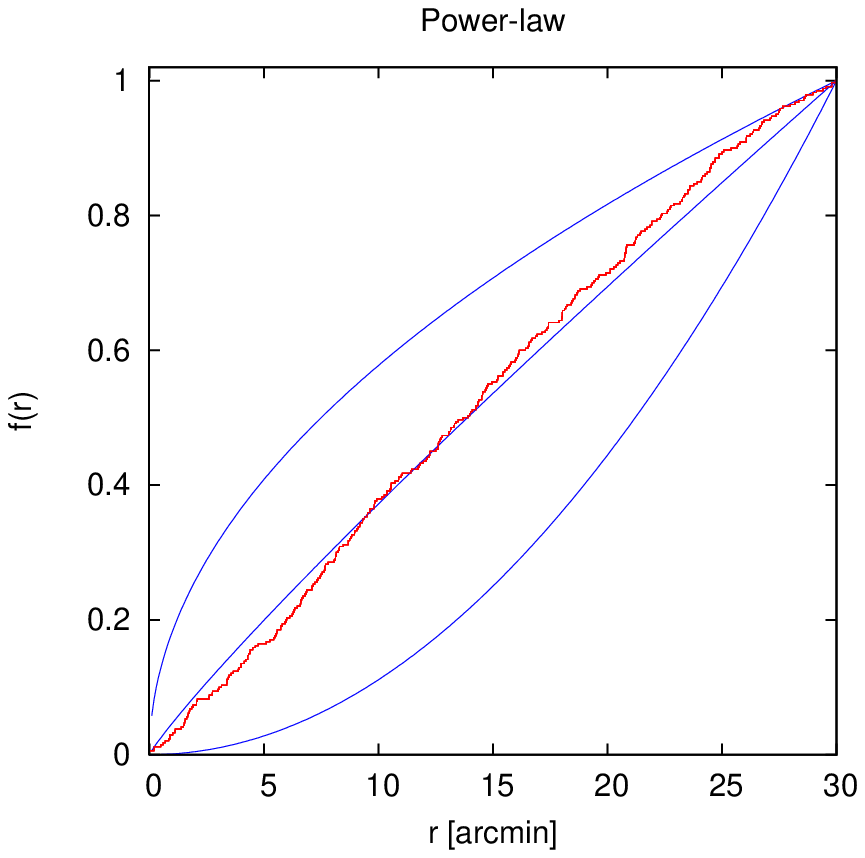}
\includegraphics[width=0.57\textwidth]{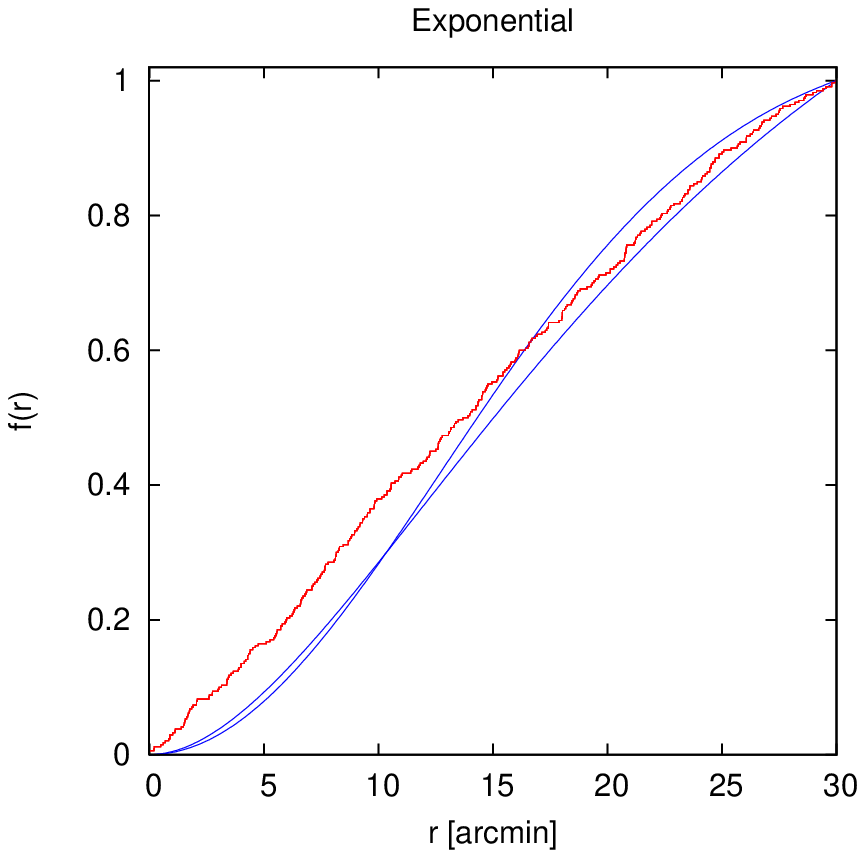}
\includegraphics[width=0.57\textwidth]{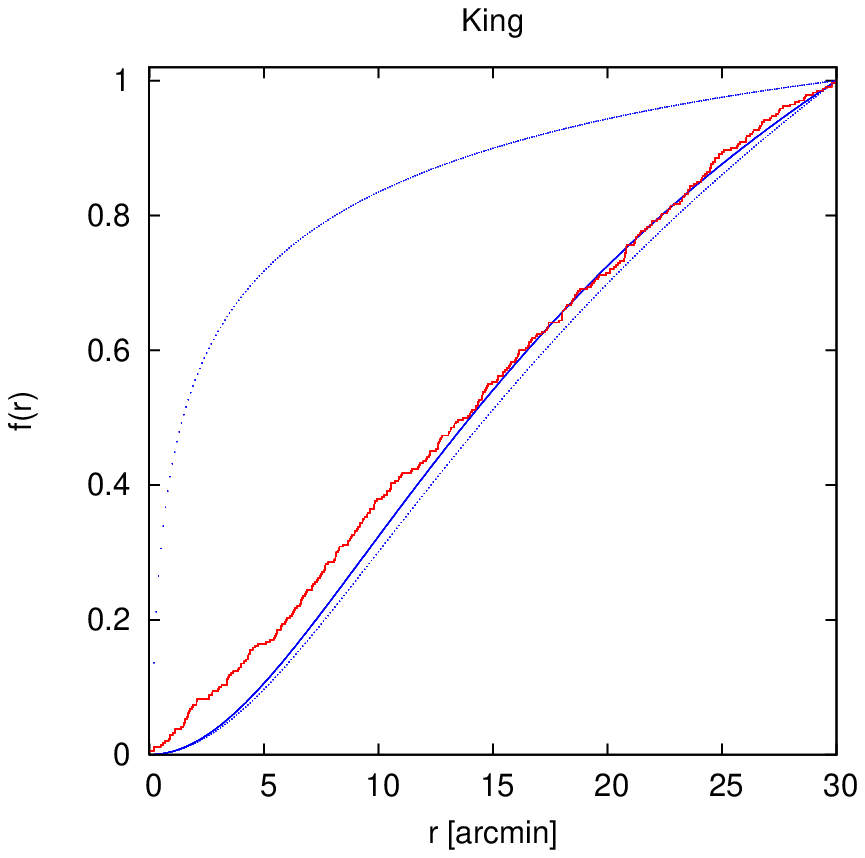}
\caption{Normalised cumulative number of $\sigma$~Orionis objects, $f^*(r)$, in
(red) steps plot style and theoretical distribution in (blue) solid line.
{\em Top panel:} power-law distributions for $f(r) \propto r^{1/2},
~ r^{0.9}, ~ r^2$ (from top to bottom);
{\em middle panel:} exponential distributions for $\sigma (r) \propto
e^{-\epsilon r}$ ($\epsilon$ = 1/12\,arcmin$^{-1}$) and $e^{-\epsilon r^2}$
($\epsilon$ = 1/18$^2$\,arcmin$^{-2}$);
{\em bottom panel:} King distributions for the core, limit (dotted) and
overall radial profiles ($r_c$ = 10\,arcmin, $r_t$ = 200\,arcmin).}
\label{ofrrs}
\end{figure}
%

\begin{figure}
\centering
\includegraphics[width=0.57\textwidth]{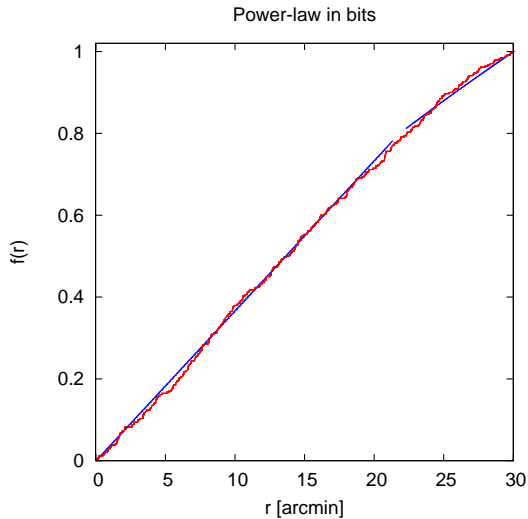}
\caption{Same as Fig.~\ref{ofrrs} but for a composite theoretical distribution 
with $f(r) = 1.1 \left(\frac{r}{r_{\rm max}} \right)^{1.0}$ for $r <$ 21\,arcsec
and $f(r) = 1.0 \left(\frac{r}{r_{\rm max}} \right)^{0.7}$ for $r >$ 23\,arcsec.}
\label{ofrr}
\end{figure}

Fig.~\ref{ofrrs} illustrates the fits of $f(r)$ to $f^*(r)$ to
evaluate the most suitable expression for $\sigma(r)$. 
The best match for a simple power-law density is aquired for
$\delta$ = 0.9.  
Power-laws with $\delta \gg 1$ and $\delta \ll 1$ provide inaccurate fits.
Likewise, the exponential profiles cannot predict the large actual surface
density close to the cluster centre (the binned surface density profile in
Fig.~\ref{orhist}, when plotted in logarithmic scale, also shows that the
innermost bin deviates from the exponential profile).
The overall King profile has the same problem.
I~performed intensive computations, not shown here, to cover the $(r_t,r_c)$
parameter space of the King profile.
No clear minimum of the $\chi^2$ exists for the $\sigma$~Orionis radial
distribution when fitted to the King empirical density law.
The best solutions were found for all the combinations that satisfy $r_c$ =
8--12\,arcmin and $r_t \gg r_c$. 
The excesses of light at {\em large} radii of young massive clusters with
respect to King (and Elson, Fall \& Freeman 1987) profile(s), attributed to gas
expulsion by Goodwin \& Bastian (2006), cannot explain the poor fitting for the
King profile at {\em small} radii in $\sigma$~Orionis. 

The best general fit is obtained for a composite power-law, as shown in
Fig.~\ref{ofrr}.  
The cumulative number of $\sigma$~Orionis objects grows proportional to $r$
(i.e. $\delta$ = 1.0) up to $\sim$20\,arcmin.
This size translates into a physical radius of $\sim$2\,pc.
At larger separations, $f(r)$ increses with a lower slope, indicating an
exponent $\delta \approx 0.7$.
In reality, at such separations, an exponential or a limit King [$\sigma_t(r)$]
profile would also fit the data.
From the extrapolation of the $f(r) \propto r$ law up to the radius of
30\,arcmin, there is deficit of 30--40 objects in the outermost annulus.
As a result, the $\sigma$~Orionis cluster may be spatially described as a
central ($r \la$ 20\,arcmin), dense region --the ``core''-- and an outer ($r
\ga$ 20\,arcmin), more rarified region --the ``halo''--. 
From these data, the power-law index transition between the core and the halo is
quite smooth. 
However, the relative drop of $f(r)$ at $r \sim$ 20\,arcmin might be more or
less abrupt because of our poor knowledge of the $\sigma$~Orionis stellar
population at large separations from the cluster centre.
As an example, while more than 90\,\% of cluster members and candidates of the
{\em Mayrit} catalogue in the innermost 10\,arcmin have known features of
extreme youth, this ratio is about 50\,\% in the halo (past spectroscopic,
mid-infrared and X-ray analysis have been naturally focused on the cluster
centre and its~surroundings).
Going deeply into this subject seems to be meaningless because of the
relatively small amount of investigated cluster objects in the halo,
contamination by young neighbouring stellar populations in Orion (Jeffries
et~al. 2006; Caballero 2007a) or fore-/background stars (e.g. Caballero,
Burgasser \& Klement, in prep.), variable extinction to the northeast of the
survey area, incompleteness of the {\em Mayrit} catalogue (Caballero 2007c)
and the ``anomalous'' radial distribution discussed~next. 

The radial distribution of the cluster core is, on the contrary to the
halo, free of possible systematic errors. 
The $f(r) \propto r$ law in the core is translated to surface and volume
densities $\sigma(r) \propto r^{-1}$ and $\rho(r) \propto r^{-2}$.
Radial profile investigations carried out in other very young star-forming
regions have found similar distributions. 
Not far away from $\sigma$~Orionis, Bate, Clarke \& McCaughrean (1998) noticed
that the stars of the Orion Nebula Cluster are distributed with a core of
uniform volume density and radius $r_{\rm core}$ = 0.5\,arcmin and a volume
density profile $\rho(r) \propto r^{-2}$ at larger separations. 
Alike power-law volume densitiy distributions have been found in other
star-forming regions, like Taurus (Fuller \& Myers 1992; Ward-Thompson et~al.
1994), or low-mass cold dark molecular clouds, like {Barnard~68} (Alves, Lada \&
Lada 2001). 
Furthermore, the power-law index 2 is an {\em ``often-used initial condition for
numerical calculations of star formation''} (Burkert, Bate \& Bodenheimer 1997).
The $\rho(r) \propto r^{-2}$ distribution corresponds to a singular,
self-gravitating, (rotating) isothermal sphere. 
Finally, from fig.~5 in Cartwright \& Whitworth (2004) and the ${\mathcal
Q}$-parameter value for $\sigma$~Orionis derived in Section~\ref{clustercentre},
I~estimate that the volume density in the cluster varies as $\rho(r) \propto
r^{-1.7 \pm 0.4}$.
The --1.7$\pm$0.4 index is intermediate between those in $\rho$~Ophiuchus
(--1.2) and IC~348 (--2.2) and consistent with $\rho(r) \propto r^{-2}$.

The use of the power-law function $f(r) \propto r$, whose corresponding
$\sigma(r)$ diverges at $r$ = 0, has the drawback of an incorrect fit in the
innermost 1\,arcmin of the cluster. 
Characterising the very centre of the cluster is out of the scope of this work,
since it can be only accomplished with high spatial-resolution facilities (e.g.
adaptive optics or mid-infrared instruments -- van Loon \& Oliveira 2003;
Caballero 2005). 
Moreover, the Bate et~al. (1998)'s value of $r_{\rm core}$ can be suitable
applied to $\sigma$~Orionis.
There are only seven stars at less than 0.5\,arcmin to $\sigma$~Ori~AB
(Caballero 2007b), so the central surface density is $\sigma_0 \approx$
9\,arcmin$^{-2}$. 
This value matches well with the actual value of $\sigma(r_{\rm core})$ for
$f(r) \propto r$. 
The observational parallelism between the radial distributions of the Orion
Nebula Cluster and other star-forming regions and of $\sigma$~Orionis evidences
that the collapse of an isothermal cloud to form a star cluster might
be~universal.

\subsection{Mass-dependent radial distribution}

\begin{figure}
\centering
\includegraphics[width=0.57\textwidth]{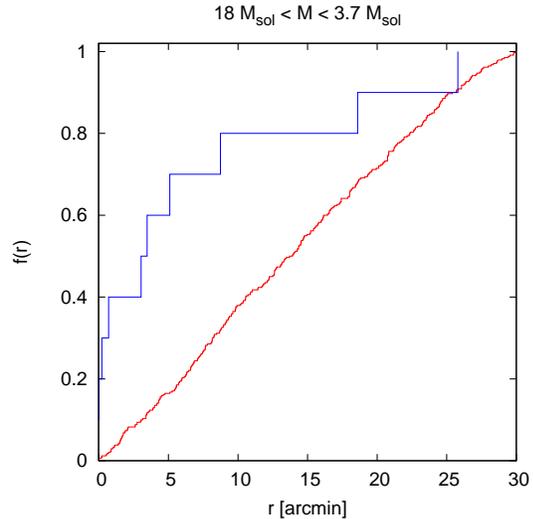}
\caption{Mass-dependent normalised cumulative number of high-mass
$\sigma$~Orionis  
stars as a function of the separation to the cluster centre.
The observed $f^*(r)$ for the 340 objects is overplotted as in
Figs.~\ref{ofrrs}~and~\ref{ofrr}.}
\label{ofrrmA}
\end{figure}
%

\begin{figure}
\centering
\includegraphics[width=0.57\textwidth]{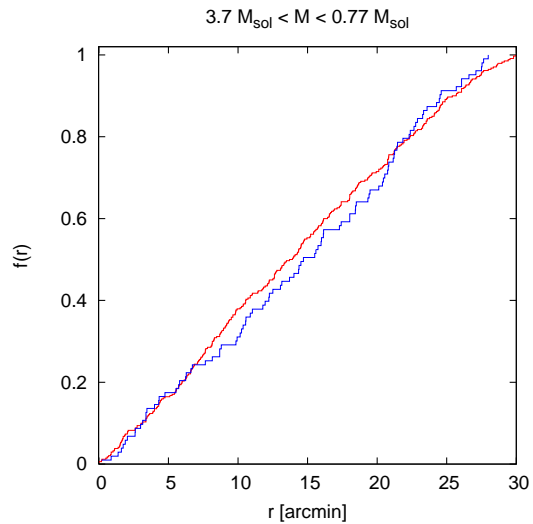}
\caption{Same as Fig.~\ref{ofrrmA}, but for intermediate-mass stars.}
\label{ofrrmB}
\end{figure}
%

\begin{figure}
\centering
\includegraphics[width=0.57\textwidth]{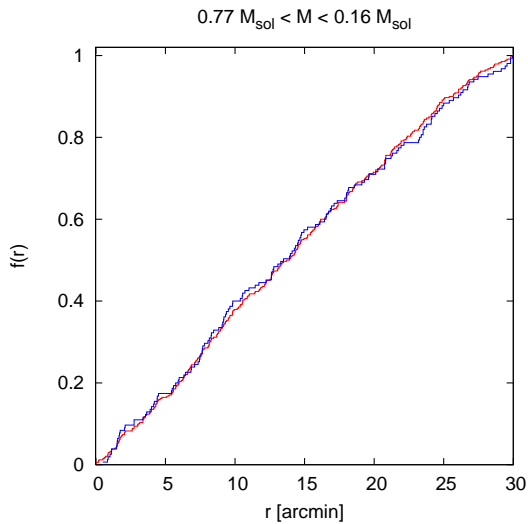}
\caption{Same as Fig.~\ref{ofrrmA}, but for low-mass stars.}
\label{ofrrmC}
\end{figure}
%

\begin{figure}
\centering
\includegraphics[width=0.57\textwidth]{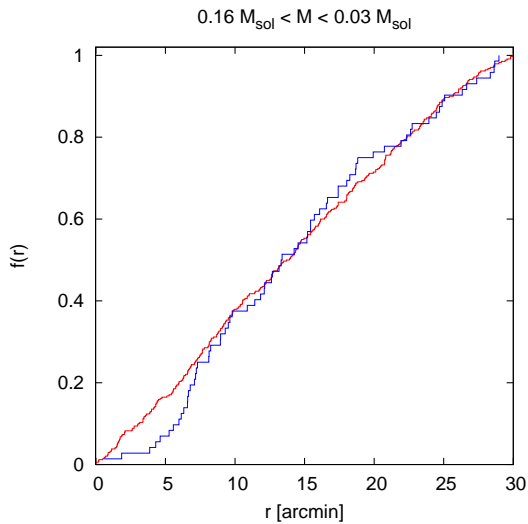}
\caption{Same as Fig.~\ref{ofrrmA}, but for very low-mass stars and
brown~dwarfs.
Note the deficit of very low-mass cluster member and candidates at $r \la$
4\,arcmin and the inclined raise at $r =$ 6--7\,arcmin} 
\label{ofrrmD}
\end{figure}

I have also investigated the radial profile of $\sigma$~Orionis cluster
members and candidates for different mass intervals.
I~have separated the 340 stars and brown dwarfs in four mass groups that are
equally spaced in logarithmic scale.
The boundaries between the groups are at about 3.7, 0.77 and 0.16\,$M_\odot$.
The groups contain, from the most to the least massive, 10, 103, 155 and 72
elements.
Figs.~\ref{ofrrmA}~to~\ref{ofrrmD} show the normalised cumulative number of
objects for the four groups compared to the average distribution.
It is manifest that the ten stars more massive than 3.7\,$M_\odot$ depart from
the general trend $f(r) \propto r$, while the brown dwarfs and stars below this
mass roughly follow it. 
The massive stars seem to obey the King profile close to the cluster centre
(with a surface density $\sigma_c(r) = \frac{f_0}{1+(r/r_c)^2}$) rather than a
power-law, whose exponent should be as low as $\delta \sim$ 0.1.
Eight of the ten most massive $\sigma$~Orionis stars are, besides, within the
10\,arcmin-radius circle.
This agglomeration of early-type stars towards the cluster centre is typical in
other very young star-forming regions, such as the Orion Nebula Cluster, whose
centre is defined by the OB-type stars of the {$\theta^1$~Ori} multiple
system. 
This result shows again the resemblance between both Orion~clusters.

There is no other characteristic feature in the mass-dependent distribitution in
Figs.~\ref{ofrrmA}~to~\ref{ofrrmD}, except for a remarkable deficit of very
low-mass stars and high-mass brown dwarfs (0.16\,$M_\odot$ $\le M \le$
0.03\,$M_\odot$) in the innermost 4\,arcmin together with a steep raise of
$f^*(r)$ at 6--7\,arcmin (Fig.~\ref{ofrrmD}). 
On the one hand, there are only two representatives of this mass interval within
4\,arcmin: Mayrit~36273 and the Class I object Mayrit~111208. 
Both of them are the faintest sources with $I$-band photometry identified in the
near-infrared/optical/X-ray survey in the centre of
$\sigma$~Orionis by Caballero (2007b).
On the other hand, there are almost 20 low-mass stars and brown dwarfs in the
narrow annulus 5--8\,arcmin.
An 0.16\,$M_\odot$-mass object in the cluster, in the upper limit of low-mass
interval, has typical magnitudes $I \sim$ 16.0\,mag, $J \sim$ 14.0\,mag.
These values are far brighter than the DENIS and 2MASS completenesses, even in
the innermost region affected by the glare of the multiple system $\sigma$~Ori.
Caballero (2007b) failed to confirm or refute this absence of very low-mass
stars and high-mass brown dwarfs.
The raise of $f^*(r)$ at 6--7\,arcmin indicates, on the contrary, a larger
density of very low-mass objects in this annulus.
One can think of many ways which could give rise to the deficit of low-mass
objects at small radii and excess at intermediate radii.
Firstly, low-mass objects could actually form in the 6--7\,arcmin annulus and
not in the  inner regions (maybe core masses were higher in the centre, or
competitive accretion caused central brown dwarfs to grow into stars). 
Alternatively, they could form in the cluster centre, but were ejected via
dynamical interactions with the massive stars.
Low-mass objects do not have enough energy, however, to move further away from
the deep $\sigma$~Ori gravity well of $M >$50\,$M_\odot$.
The deficit-excess needs to be explained by theory, but this particular set
of observations does not give any indication of a preferred formation scenario
[see Whitworth et~al. (2007) for a review on the theory of formation of brown
dwarfs and very low-mass stars].
Further and innovative observations, able to avoid the extense glare of the OB
system, are required to determine if the peculiar distribution of very low-mass
objects in the cluster centre are due to an observational bias or to an actual
consequence of the formation mechanism. 
Some observational efforts on this topic have been carried out by Caballero
(2005, 2007b) and Sherry, Walter \& Wolk~(2005).

\subsection{Azimuthal asymmetry}
\label{azimuthalasymmetry}

\begin{figure}
\centering
\includegraphics[width=0.57\textwidth]{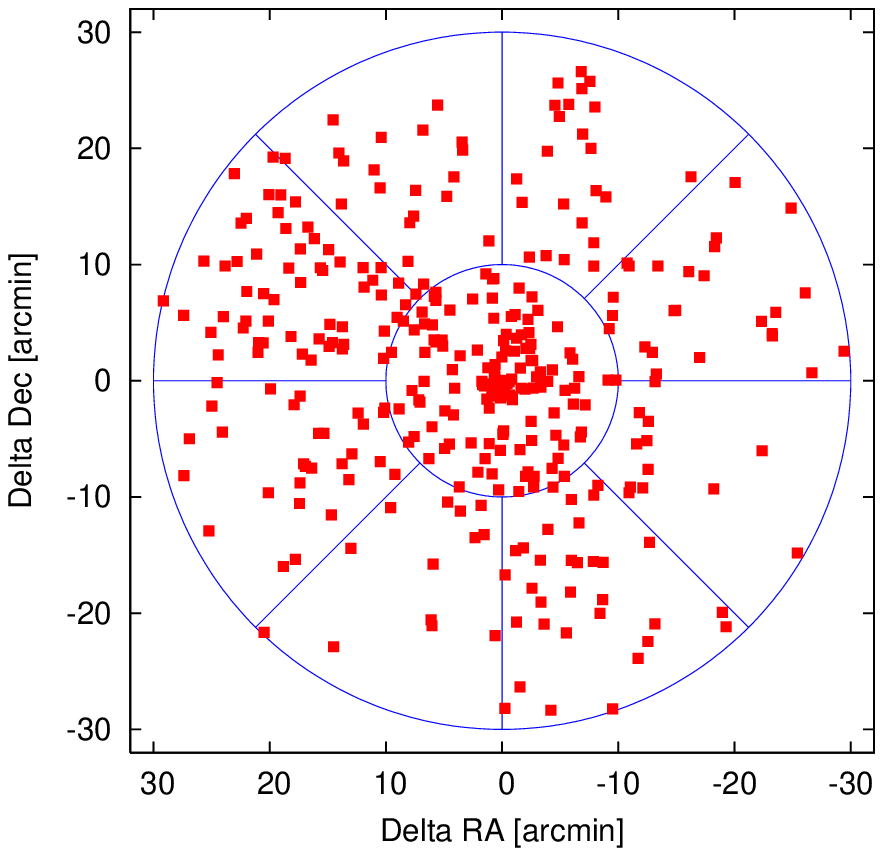}
\includegraphics[width=0.57\textwidth]{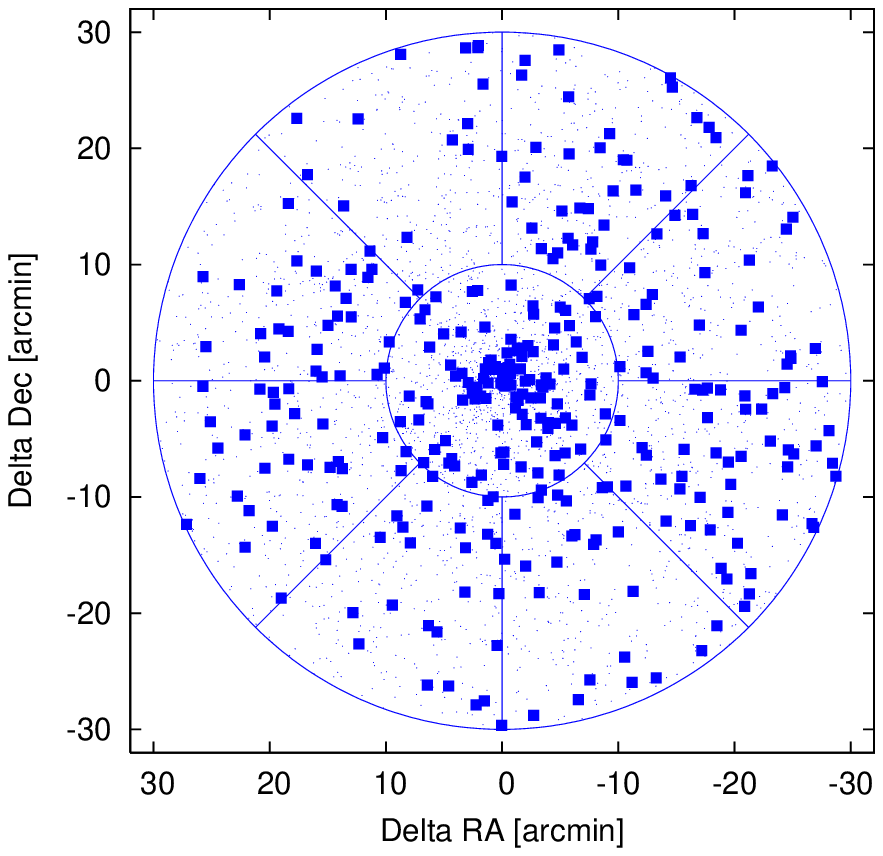}
\caption{Actual (top) and simulated (bottom) spatial distributions of
$\sigma$~Orionis objects in the survey area.
The coordinate origin is in $\sigma$~Ori~A.
The bottom window shows 10 Monte Carlo simulations following $f(r) =
\left(\frac{r}{r_{\rm max}} \right)^{1.0}$.
One distribution is marked with small filled squares, the remaining ones are
marked with points. 
The sizes of the small and big solid circles are 10 and 30\,arcmin.
The eight equal-size segments of annuli used for the computation of the
asymmetry factor, $\zeta$, are also indicated.}
\label{orades}
\end{figure}

Top window in Fig.~\ref{orades} shows that the distribution of confirmed
cluster members and candidates is not whole radially symmetric, with an evident
lower density to the west of $\sigma$~Ori~AB with respect to the east.
An elongated subclustering is manifest to the east-northeast of the cluster
centre, just in the direction to the Horsehead Nebula.
On the contrary, B\'ejar et~al. (2004) derived that the variation of the radial
distribution of their very-low mass stars and brown dwarfs over their best
exponential law fit was Poissonian, implying no evidence of subclustering.
The object sample presented in this paper surpasses B\'ejar et~al.'s one and
allows to corroborate or invalidate their statement.

%

I~have looked for an azimuthal asymmetry in the $\sigma$~Orionis cluster in
several steps.
First, I~have generated 1000 simulated distributions\footnote{In the general
power-law distribution case with index $\delta$, if $\bmath{r}$ is a vector of
length $N_{\rm max}$ of uniformly distributed pseudo-random radii between $r$ =
0 and $r_{\rm max}$, then the distribution of the vector $\bmath{r}^{1/\delta}$
follows $f(r) = \left(\frac{r}{r_{\rm max}} \right)^\delta$.} following the
power law $f(r) = \left(\frac{r}{r_{\rm max}} \right)^{1.0}$.
Ten of them (one is highlighted) are shown in the bottom window in
Fig.~\ref{orades}.
The resemblance with the actual distribution, in the top window of the Figure,
is evident. 
Second, I~have divided the survey area in nine regions: a central area,
10-arcmin wide, where the asymmetry is difficult to quantify, and eight segments
of the annulus in the radius interval 10\,arcmin $< r <$ 30\,arcmin. 
Third, for each simulated distribution, I~have computed its corresponding
asymmetry factor, $\zeta$, defined by:

\begin{equation}
\zeta = \frac{(N_1+N_2)-(N_7+N_8)}{N_{\rm max}},
\end{equation}

\noindent where $N_1$ and $N_2$ are the numbers of objects of the two most
populated segments of annulus, and $N_7$ and $N_8$ are those of the less
populated. 
The mean and the standard deviation of the 1000 computed asymmetry factors are
$\overline{\zeta} = 0.0698$, $\sigma_\zeta = 0.0189$. 
The maximum and minimum asymmetry factors are $\zeta_{\rm max}$ = 0.1382 and
$\zeta_{\rm min}$ = 0.0176.
The four values ($\overline{\zeta}$, $\sigma_\zeta$, $\zeta_{\rm max}$ and
$\zeta_{\rm min}$) are faultless compatible with the Poissonian errors within
each segment of annulus.
Finally, I~have measured the asymmetry factor for the actual $\sigma$~Orionis
distribution: $\zeta^*$ = 0.1853.
This value deviates 6.1 times the $\sigma_\zeta$ from the $\overline{\zeta}$ and
is significatively larger than the $\zeta_{\rm max}$ among 1000 Monte Carlo
simulations. 
Considering as a first order of approximation that the values $\zeta_i$ ($i$ =
1...1000) are distributed following a standard normal distribution with
parameters $\overline{\zeta}$ and $\sigma_\zeta$, then there is a probability $p
= 1 - {\rm erf}(6.1/2^{1/2}) \approx 10^{-9}$ (where ``erf'' is the error
function) that the actual value $\zeta^*$ follows such distribution.
Even accounting for generous systematic errors or biases, it is highly probable
that the radial distribution of objects at more than 10\,arcmin from the
centre of $\sigma$~Orionis is azimuthally~asymmetric.
 
The most populated segments of annulus of the actual distribution are, counting
anti-clockwise from 12 hours, the second (coinciding with the elongated
subclustering in the direction to the Horsehead Nebula) and the fifth ones.
Both two segments and the cluster centre spatially coincide with a filamentary
region of maximum emission at the 12\,$\mu$m {\em IRAS} passband (see fig.~2 in 
Oliveira \& van Loon 2004). 
This ``warmer'' region has not been identified in works based on recent
observations with IRAC and MIPS onboard {\em Spitzer} (e.g. Hern\'andez et~al.
2007) or with the Spatial Infrared Imaging Telescope on the Midcourse Space
Experiment satellite (Kraemer et~al. 2003).
From my data, I cannot postulate whether the largest surface density of
$\sigma$~Orionis objects originally arises from an hypothetical larger density
of warm dust in the region or, inversely, the filamentary region is a
consequence of both the low spatial resolution imaging capabilities of {\em
IRAS} and the largest surface density of objects (i.e. the red $\sigma$~Orionis
objects, many with mid-infrared excesses due to discs --Oliveira et~al.
2006; Hern\'andez et~al. 2007; Caballero et~al. 2007--, generate a smooth
background at 12\,$\mu$m that {\em IRAS} was not able to resolve).

The accumulation of stars and brown dwarfs in a filamentary pattern in
$\sigma$~Orionis strongly supports some star formation scenarios of collapse and
fragmentation of a large-scale turbulent molecular cloud, especially those that
predict burst of star formation in filamentary gas ({e.g.} Bate, Bonnell \&
Bromm~2003).
It is stimulating to notice that these simulations assumed the contraction of an
isothermal, spherical molecular cloud with $\rho(r) \propto r^{-2}$ (see
Section~\ref{surfacedensity}).
The filamentary accumulation in $\sigma$~Orionis is, however,
peculiar, because no similar arrangements have been found in other star-forming
regions. 
For example, G\'omez et~al. (1993) and Larson (1995) found that the
subclustering in {Taurus-Auriga} is in the form of star clumps of
$\sim$15 components, while Bate et~al. (1998) showed that in the Orion Nebula
Cluster there is no subclustering at all.
It is obvious that further investigations are needed; percolation or two-point
correlation function of stars are different approaches that can be~used.


\section{Summary}

The $\sim$3\,Ma-old $\sigma$~Orionis cluster is a perfect laboratory of
star formation. 
I~have investigated the radial distribution of 340 cluster members and
candidates in a 30\,arcmin-radius area centred on $\sigma$~Ori~AB, taken from
Caballero (2007c). 
The analysis has covered a mass interval from the 18+12\,$M_\odot$ of
$\sigma$~Ori~AB to the $\sim$0.03\,$M_\odot$ of the faintest brown dwarf
detectable by DENIS. 
The cluster shows a clear radial density gradient, quantified by the
${\mathcal Q}$-parameter, that accounts for the mean separation between members
and the Euclidean minimum spanning tree of the cluster.
I~have calculated the functional relations between normalised cumulative numbers
of objects counting from the cluster centre, $f(r)$, and surface densities,
$\sigma(r)$.
Cumulative distribution functions as these avoid many problems associated
with binning.
Among the studied radial (power-law, exponential and King) profiles, the best
fit is for a composite power-law distribution of cluster members with a core and
a rarified halo.
The core extends up to $\sim$20\,arcmin from the cluster centre and is nicely
modelled by a surface density $\sigma(r) \propto r^{-1}$, that corresponds to a
volume density $\rho(r) \propto r^{-2}$.
This volume density matches, in its turn, the radial profile in a cluster formed
from the collapse of a self-gravitating, isothermal sphere. 
The most massive $\sigma$~Orionis stars deviate, however, from the general trend
and are much more concentrated towards the cluster centre.
There is also an apparent deficit of very low-mass stars and high-mass brown
dwarfs (0.16\,$M_\odot \ga M \ga$ 0.035\,$M_\odot$) in the innermost 4\,arcmin
and an excess in the annulus at 6--7\,arcmin to the central Trapezium-like
system. 
Last, there is a significant azimuthal asymmetry due to a filament-shape
overdensity of objects that connects the cluster centre with a part of the
Horsehead Nebula.  
This discovery supports the formation scenarios that predict burst of star
formation in filamentary~gas.

\section*{Acknowledgments}

I thank the anonymous referee for helpful comments.
J.A.C. was formerly an Alexander von Humboldt Fellow at the MPIA, and is
currently an Investigador Juan de la Cierva at the UCM.
Partial financial support was provided by the Universidad Complutense de Madrid
and the Spanish Ministerio Educaci\'on y Ciencia under grant 
AyA2005--02750 of the Programa Nacional de Astronom\'{\i}a y Astrof\'{\i}sica
and by the Comunidad Aut\'onoma de Madrid under PRICIT project S--0505/ESP--0237
(AstroCAM).

\bsp

\label{lastpage}

\end{document}